\documentclass{emulateapj}

\usepackage{hyperref,ifthen,psfig,graphicx}
\usepackage{epstopdf}
\usepackage{soul,color,amsmath}
\usepackage{ifthen,psfig,graphicx}
\usepackage{epstopdf,aas_macros}
\usepackage{soul,color,amsmath}
\newcommand{\myemail}{\email{wpmaksym@ua.edu}}

\def\mathnew{\mathsurround=0pt}
\def\simov#1#2{\lower 2.5pt\vbox{\baselineskip0pt \lineskip-.5pt
\ialign{$\mathnew#1\hfil##\hfil$\crcr#2\crcr\sim\crcr}}}
\def\simless{\mathrel{\mathpalette\simov <}}
\def\simgreat{\mathrel{\mathpalette\simov >}}
\newcommand{\MeV}{Me\kern-0.11em V}
\newcommand{\keV}{ke\kern-0.11em V}

\newcommand{\cha}{{\it Chandra\/}}
\newcommand{\xmm}{{\it XMM-Newton\/}}

\newcommand{\ecmss}{erg~cm$^{-2}$ s$^{-1}$}
\newcommand{\es}{erg~s$^{-1}$}
\newcommand{\omegam}{$\Omega_{m,0}$\/}
\newcommand{\omegal}{$\Omega_{\Lambda,0}$\/}

\newcommand{\Mbh}{\ensuremath{M_{\bullet}}}
\newcommand{\Msun}{\ensuremath{{\rm M}_{\odot}}}

\newcommand{\tdr}{\ensuremath{\gamma_{D}}}
\newcommand{\tdru}{\ensuremath{\rm{galaxy}^{-1}\rm{year}^{-1}}}
\makeatletter
\newcommand\iont[2]{{#1$\;${\small\expandafter\@slowromancap\romannumeral #2@\relax}}}
\makeatother
\makeatletter
\newcommand{\raisemath}[1]{\mathpalette{\raisem@th{#1}}}
\newcommand{\raisem@th}[3]{\raisebox{#1}{$#2#3$}}
\makeatother

\slugcomment{Accepted by ApJL, 5 July 2014}

\shorttitle{RBS 1032: A TDE in a Dwarf Galaxy?}
\shortauthors{Maksym et al.}

\begin{document}


\title{RBS 1032: A Tidal Disruption Event in Another Dwarf Galaxy?}


\author{W. Peter Maksym}
\affil{University of Alabama, Department of Physics and Astronomy, Tuscaloosa AL 35487}
\myemail

\author{Dacheng Lin}
\affil{Space Science Center, University of New Hampshire, Durham, NH 03824}
\affil{University of Alabama, Department of Physics and Astronomy, Tuscaloosa AL 35487}

\and

\author{Jimmy A. Irwin}
\affil{University of Alabama, Department of Physics and Astronomy, Tuscaloosa AL 35487}



\begin{abstract}
\object{RBS 1032} is a supersoft ($\Gamma\sim5$), luminous ($\sim10^{43}\;$\es) {\it ROSAT} PSPC source which has been associated with an inactive dwarf galaxy at $z=0.026$, \object{SDSS J114726.69+494257.8}.  We have analyzed an \xmm\ observation which confirms that RBS 1032 is indeed associated with the dwarf galaxy.  Moreover, RBS 1032 has undergone a factor of $\sim100-300$ decay since November 1990.  This variability suggests that RBS 1032 may not be a steadily accreting intermediate-mass black hole, but rather an accretion flare from the tidal disruption of a star by the central black hole (which may or may not be intermediate-mass).  We suggest that additional tidal disruption events may remain unidentified in archival {\it ROSAT} data, such that disruption rate estimates based upon {\it ROSAT} All-Sky Survey data may need reconsideration.

\end{abstract}




\section{Introduction}

Most (if not all) galaxies are thought to host massive ($\Mbh\simgreat10^5\;\Msun$) black holes (MBHs) at their nuclei.  An important consequence is that occasionally a star may pass within the tidal disruption radius ($R_T$) of the MBH and be disrupted.  The tidal forces overwhelm the star's self-gravity, and the star is ripped apart in a tidal disruption event \citep[TDE;][]{Hills75,Rees88}.  The bound debris falls back onto the MBH, generating an accretion-powered flare whose evolution is (to first order) governed by the approximately Keplerian orbits and the energy spread of the debris.  The flare itself is thought to typically be most luminous in X-rays and ultraviolet \citep[e.g.,][]{Ulmer99} but may also produce a relativistic jet which emits hard X-rays and gamma-rays \citep{Bloom11,Burrows11}.  The rate at which TDEs occur (\tdr) is furthermore important to models of galaxy formation and evolution, as \tdr\ is a function of the MBH population distribution and the stellar population in the central cluster of a given galactic nucleus.

Recent wide-field time domain surveys have proven particularly effective at detailing TDE light curves during their rise and decline \citep[e.g.][]{Gezari12}.  Since soft X-rays are likely to contain the bulk of the bolometric luminosity, however, numerous X-ray searches at $kT\simless2\;$keV have long proven productive in finding TDEs (e.g. \citealt{Maksym10,Maksym13,Maksym14,Lin11,Irwin10,Saxton12,Esquej08}).  The earliest confident TDEs were detected using {\it ROSAT}  \citep{BKD96,KB99}, with one of the most convincing early cases being RX J1242.6-1119A \citep{KG99,Komossa04}.  By extension, one of the most-cited observational determinations of \tdr\ was conducted by \cite{Donley02} using data from the {\it ROSAT} All-Sky Survey \citep[{\it RASS};][]{RASS}.  {\it ROSAT} continues to be useful in the identification of new TDEs, not only as a critical indicator of pre-flare upper limits, but also in the identification of at least one new flare whose extreme variability was undetected by {\it ROSAT}, but which became evident through later X-ray observations \citep{Cappelluti09}.

RBS 1032 was a bright ($F_X[\rm{0.1-2.4\;keV}]\simgreat10^{-12}\;$\ecmss), ultrasoft X-ray source identified in {\it RASS} \citep{RBS1}.  The optical counterpart was initially suggested to be a star by \cite{Zickgraf03}.  Later work by \cite{Ghosh06} using the Sloan Digital Sky Survey \citep{SDSSdr10}, the Russian-Turkish 1.5-m telescope in Anatalya, Turkey, and the 6-m telescope of the Special Astrophysical Observatory in Russia, however, showed the most likely optical counterpart to be a dwarf galaxy at $z\sim0.026$ without emission lines, SDSS J114726.69+494257.8.  They considered, and rejected as unlikely, numerous possible explanations for the X-ray emission.  They also suggested that the system may be a binary with an intermediate-mass black hole (IMBH) as the primary component and a star, possibly a white dwarf, as the secondary.

We have investigated the most recent X-ray observation of RBS 1032 taken using \xmm, and suggest an alternative explanation that RBS 1032 is certainly a transient source, and is a strong candidate for a TDE.  If so, this adds to the list of TDEs reported in dwarf galaxies which may indicate the presence of IMBHs in dwarf galaxies, including WINGS J1348 in Abell 1795 (\citealt{Maksym13,Maksym14,Donato14}) and GRB 060218 \citep{Shcherbakov13}.  Such IMBHs are important clues to the formation of the first black holes \citep[and references therein]{Maksym13}.  Furthermore, we note that the discovery of new TDEs in archival {\it ROSAT} data has implications for \tdr\ as determined by \cite{Donley02}.  We have conducted our analysis and arrived at this conclusion independently from the recent paper by \cite{KS14}, who count RBS 1032 among their TDE candidates.  We find that, with minor differences, our analysis is largely in agreement with theirs.

Throughout this paper, we adopt concordant cosmological parameters\footnote{Distances are calculated according to http://www.astro.ucla.edu/~wright/CosmoCalc.html} of
$H_0=70\ $km$^{-1}$ sec$^{-1}$ Mpc$^{-1}$, \omegam=0.3 and \omegal=0.7. All coordinates are J2000.  
%, and calculate distances using \cite{Wright06}. All coordinates are J2000.  

\section{Observations and Data}

\subsection{Previous {\it ROSAT} Work}

RBS 1032 was detected by {\it ROSAT} as part of RASS on November 5, 1990, and was re-observed with pointed PSPC observations on 1992 December 7 and 1994 June 05.  For these epochs, \cite{Ghosh06} determined an unabsorbed $F_X(\rm{0.1-2.4\;keV})=[6.0,2.3,1.1]\times10^{-12}\;$\ecmss, respectively\footnote{Ghosh et al. (2006) claim to have used the $1-2.4\;$keV band in their {\it ROSAT} analysis, but we believe this to be an uncorrected typographical error.  The standard broad {\it ROSAT} band is $0.1-2.4\;$keV, which is consistent with flux values calculated using WebPIMMS (http://heasarc.nasa.gov/Tools/w3pimms.html) and the \cite{Ghosh06} assumed spectral model.} assuming a blackbody with $kT_{bb}=0.055\;$keV and Galactic $N_H=1.98\times10^{20}\;\rm{cm}^{-2}$, although they also fitted the data to a variety of models, including power law and multicolor blackbody disk models.  In all epochs, they found the spectrum to be supersoft ($\Gamma_{pl}\sim5$ or $kT_{bb}\sim0.06\;$keV).

\subsection{XMM Observation}

RBS 1032 was observed by {\it XMM-Newton} on 21 November 2009 for $\sim17\;$ks (obs. id 0604020101, PI: K. Ghosh).  The observation was badly contaminated by background flaring, leaving only [1.5, 4.7, 9.8] ks of usable [PN, M1, M2] data.  Inspection of the pipeline-produced broadband image reveals a faint point source within the $\sim30\arcsec$ {\it ROSAT} PSPC PSF when compared to the location of RBS 1032 ($r\sim6.4\arcsec$ separation).  This source is also recovered by the \xmm\ pipeline, with a corresponding UVW1 ($\sim2910\;$\AA) source detected by the optical monitor pipeline at $m_{UVW1}=20.3$ ($M_{UVW1}=-15$, $\nu L_{\nu,UVW}=1.2\times10^{41}\;$\es).  This corresponds to the source \object{3XMM J114726.7+494257} $([\alpha,\delta] = [11^{h}47^{m}26^{s}.73$, $+49\degr42\arcmin57\arcsec.3$], \citealt{3XMM}; coordinates are J2000 with $1\sigma$ error of 1.45\arcsec).  This is the only such source within 30\arcsec.  The next-closest pipeline source is at $\simgreat3\arcmin$ separation.

This XMM source is $r\sim0.8\arcsec$ from SDSS J114726.69+494257.8, and within the $1\sigma$ error (1.45\arcsec\ in [$\alpha$, $\delta$]), from the 3XMM catalog position.  The next-closest SDSS source is at $r\sim6.0\arcsec$ (comparable to the \xmm\ full width at half maximum, FWHM) and 5.6 magnitudes fainter.  The UVW1 source is also coincident with SDSS J114726.69+494257.8.

%		radecerr	ep_tot				ep_ext ep_ext_ml	ep_hr1	ep_hr2	ep_hr3		ep_hr4
%INDEX PPS 	1.49	1.486e-02 ± 3.455e-03	0.00	0.00	-0.22 ± 0.23	-0.32 ± 0.30	-0.87 ± 0.43	0.07 ± 3.84

%		RA			DEC			ID				RA			DEC			det.lik.		F(0.2-12 keV)			err		HR1		HR2		HR3		HR4
%3XMM 11 47 26.73	+49 42 57.3	J114726.7+494257	176.86137	+49.71591	 3.15e+01 	1.1419e-14	5.97e-15	-0.269	-0.380	-0.795	-1.000
% Also add from webpage?
%\footnotetext{http://cxc.harvard.edu/toolkit/colden.jsp}

\subsection{Data Reduction}

\begin{figure}
\includegraphics[height=3.in,angle=-90]{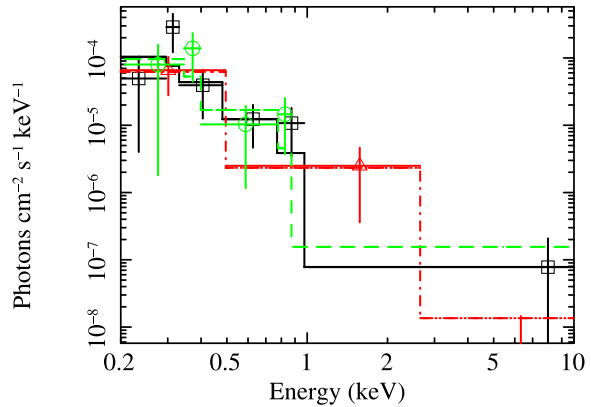} \\
\caption[]{Best-fit power law model for \xmm\ spectrum, as per Table \ref{xspec}.  Data from [PN, MOS1, MOS2] are represented as [black square, red triangle, green circle] and fit simultaneously, with model values of a given bin represented as [solid, dot-dashed, dashed] lines.  Data are fit in the $0.2-10\;\rm{keV}$ range.}
\label{xspecfig}
\end{figure}

\begin{table}
%\centering
\begin{minipage}{3.5in}
\caption{Model Fits to \xmm\ Data \label{xspec}}
\begin{tabular}{llccc}
\tableline
\tableline

Model	&	Parameter	&	Value	 & $F_X$ & $cstat/\rm{dof^\ast}$ \\
		&				&			&(0.1$-$2.4 keV)&		\\
%		&							&\multicolumn{3}{c}{($\times10^{-14}$\ecmss)}	\\
\tableline

%{\tt zpowerlw} &	$\Gamma$	& $2.88\pm0.28$	& $5.2\pm1.1$	& $5.45/5$\\
%{\tt zbbody} &	$kT_{bb}$ (keV)	& $0.11\pm0.02$	& $2.4\pm0.6$	& $6.40/5$\\
 %{\tt diskbb} &	$kT_{bb}$	(keV)	& $0.17\pm0.04$	& $2.9\pm0.7$	& $5.60/5$\\

{\tt zpowerlw} &	$\Gamma$	& $3.42\pm0.33$	& $7.4\pm3.1$	& $20.69/20$\\
{\tt zbbody} &	$kT_{bb}$ (keV)	& $0.11\pm0.01$	& $2.0\pm0.5$	& $19.62/20$\\
 {\tt diskbb} &	$kT_{bb}$	(keV)	& $0.15\pm0.02$	& $2.7\pm0.7$	& $19.23/20$\\

\tableline

\end{tabular}
\end{minipage}
%% Any table notes must follow the \end{tabular} command.\\
\\ Fits use the \cite{Cash79} statistic ($cstat$) for given degrees of freedom ($dof$).  $N_H$ is fixed and assumed at Galactic ($1.98\times10^{20}\;\rm{cm}^{-2}$) as per {\tt colden} (http://cxc.harvard.edu/toolkit/colden.jsp) and \cite{Dickey90}.  $F_X$ is in units of $10^{-14}\;$\ecmss.  $F_X$ is unabsorbed, with $1\sigma$ uncertainties.\\
\end{table}

To perform our own spectral analysis, we reduced the data for 3XMM J114726.7+494257 using {\tt XMM-SAS}\footnote{http://xmm.esac.esa.int/sas/}.  As per standard procedure, we used {\tt evselect} to filter the MOS and PN event files for periods of high background, and then to extract the spectrum for a region with $r=15\arcsec$ and $\simgreat3$ counts per bin.  We then similarly extracted a background spectrum from a nearby 33\arcsec\ region.  We then used {\tt arfgen} and {\tt rmfgen} to produce response matrix files.

We used {\tt XSPEC}\footnote{http://heasarc.gsfc.nasa.gov/xanadu/xspec/} to fit the spectra to {\tt zbbody}, {\tt zpowerlw} and {\tt diskbb} models, in order to compare to the analysis of {\it ROSAT} data by \cite{Ghosh06}.  With only $\sim17$ net PN counts (32 total), meaningful constraint of $N_H$ is not possible.  We can, however, address the general shape of the spectrum and constrain the normalization of an assumed spectral type.  We present the results of these fits in Table \ref{xspec} and Figure \ref{xspecfig}.  
%In our fits, we use \cite{cash79} statistics and assume a galactic value for  neutral hydrogen ($N_H$), as per {\small COLDEN}\footnote{http://cxc.harvard.edu/toolkit/colden.jsp} extrapolations from \cite{Dickey90}.  

\section{Light Curve Analysis}

\begin{figure*}
\centering$
\hspace{-0.2in}
\begin{array}{l}
\includegraphics[width=3.3in,angle=0]{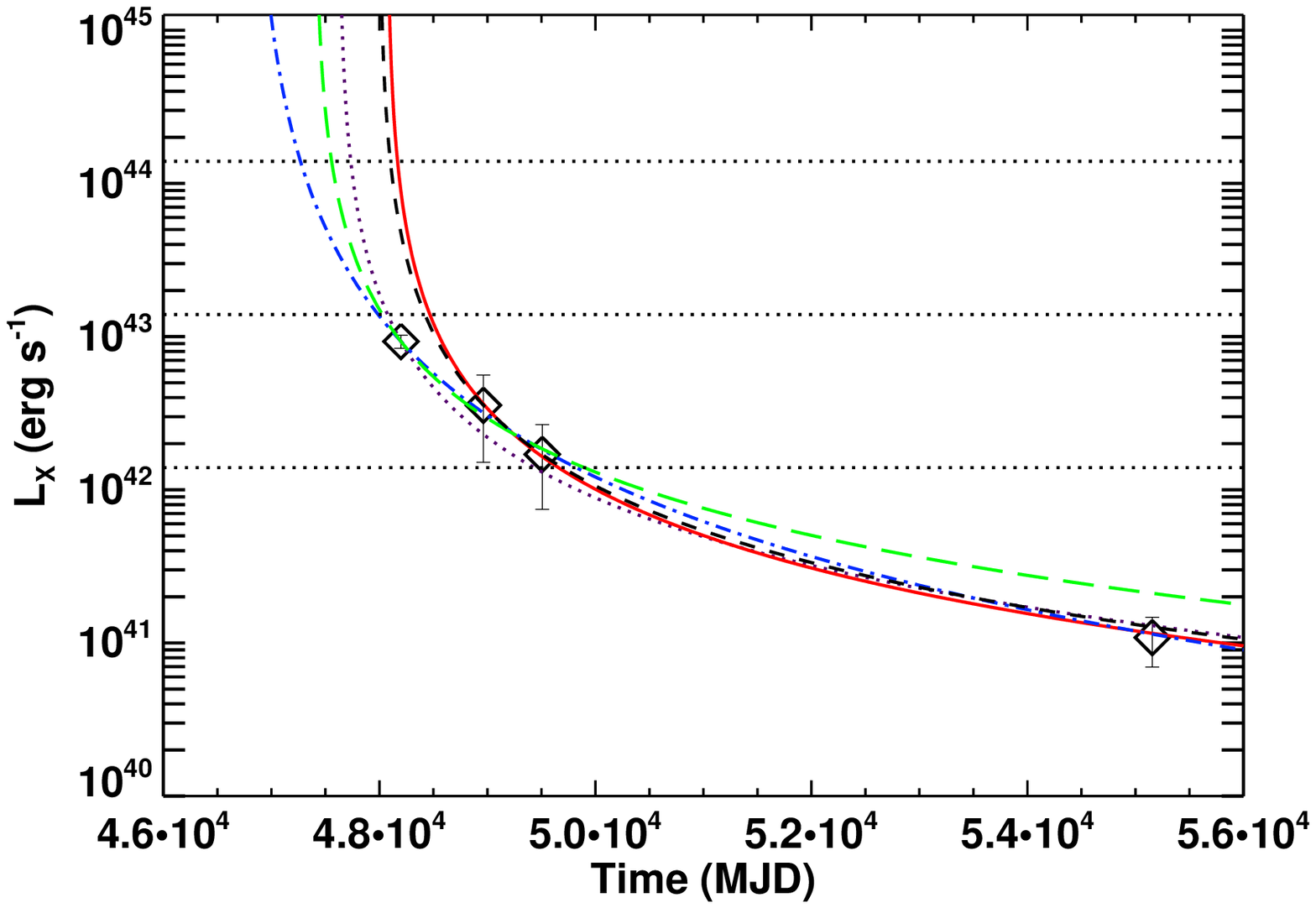} 
\hspace{0.2in}
\includegraphics[width=3.2in,height=2.2in,angle=0,clip=true, trim=0cm 0cm 8.45cm 0cm]{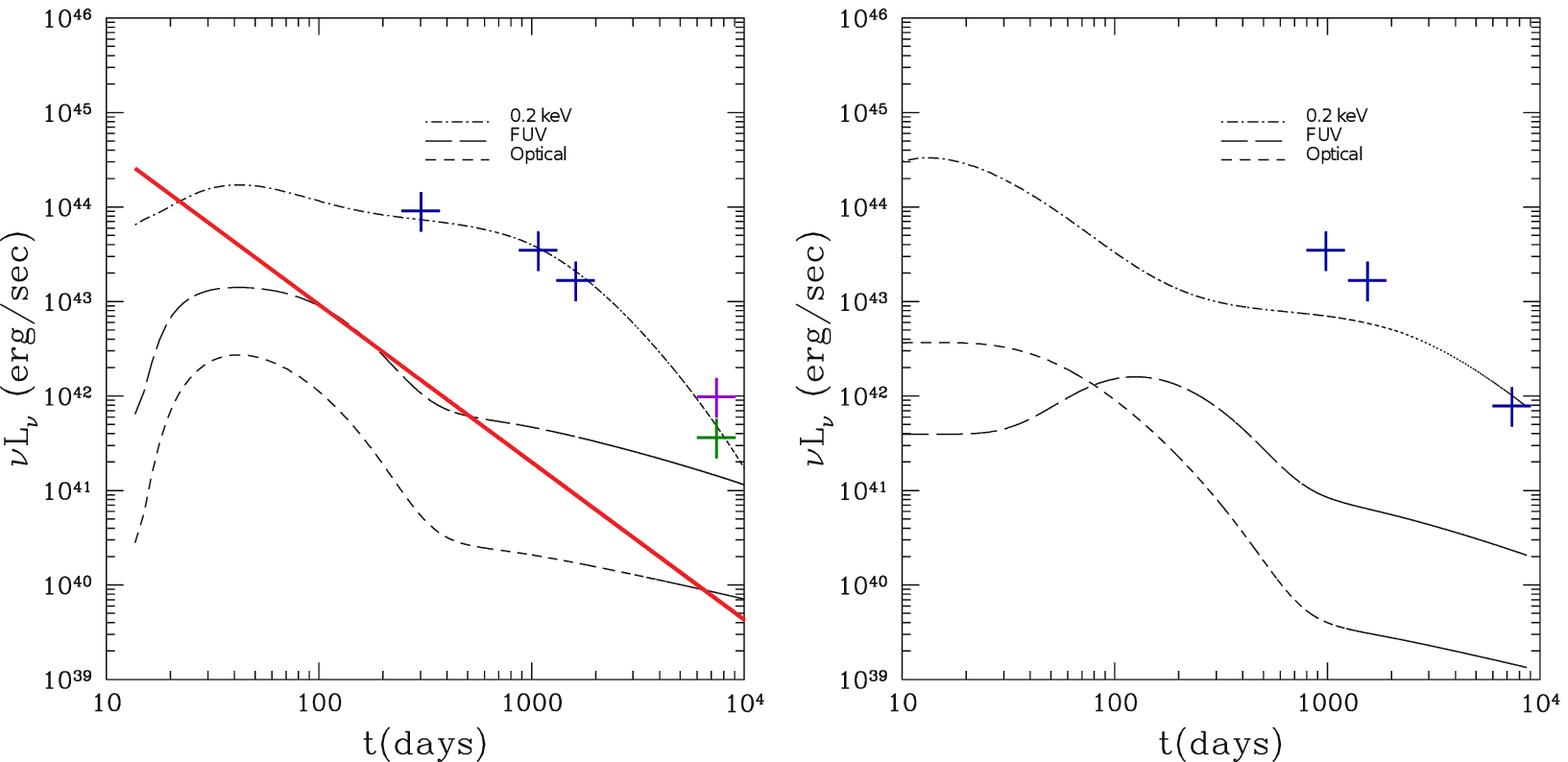} \\
\end{array}$
\caption[]{Left: Light curve and models for RBS 1032.  Except for free $n$ (blue dot-dash; $t_0=\rm{MJD}\;46806$), all curves are $t^{-5/3}$.  Other models use a subset of observed points, to emphasize possible time-dependent deviations from the basic model, and the effect on the inferred pericenter date.  Models are as follows: purple dotted (all; $t_0=\rm{MJD}\;47621$), green long-dash (1,2,3;  $t_0=\rm{MJD}\;47400$), dashed black (2,3;  $t_0=\rm{MJD}\;47995$) and solid red (2,3,4;  $t_0=\rm{MJD}\;48067$).  Horizontal dotted lines (from bottom to top) correspond to the Eddington luminosity for $\Mbh=[10^4,10^5,10^6]\;\Msun$.  Right: Figure 7 from ``Multiband light curves of tidal disruption events" by \cite{LR11}, modified to show RBS 1032 {\it ROSAT} data points (blue crosses), and the best \xmm\ {\tt zpowerlw} (purple) and {\tt zbbody} (green) fits.  Curves indicate predicted band-specific luminosities for $\Mbh=10^6\;\Msun$ for a solar-mass star with pericenter $R_T$.  The red line indicates $t^{-5/3}$ decay.  Data are scaled such that $\nu L_\nu(0.2\;\rm{keV; model})/L_X(0.1-2.4\;\rm{keV}; data)\sim9.1$ and assume $t_0=\rm{MJD}\;47900$.\\}
\label{lcfig}
\end{figure*}
%\FloatBarrier

Depending upon the spectral model, $F_X\rm(0.1-2.4\;keV)$ decreases by a factor of $\sim100-300$ from 1990 to 2009.  Given the strong variability, the extreme luminosity ($L_X>L_{Edd}$ for $\Mbh\sim10^5\;\Msun$) and the supersoft spectrum, we must therefore consider the possibility that RBS 1032 is a TDE.  At a very basic level, we can examine the X-ray light curve to compare against the assumption that the X-ray luminosity tracks the accretion rate $\dot{M}$, for which $L_X\propto(t-t_0)^{-5/3}$ (to first order), where $t_0$ is the time of stellar pericenter.

We plot a variety of simple model fits to the light curve in Figure \ref{lcfig}, assuming the best-fit value for {\tt zpowerlw} in 2009.  The data generally fit a simple $t^{-5/3}$ approximation, although the \xmm\ data point is below the prediction of a $t^{-5/3}$ curve which fits only the {\it ROSAT} points.  Unlike {\tt zpowerlw}, the inferred luminosities from the {\tt zbbody} and {\tt diskbb} models are below all light curve fits in Fig. \ref{lcfig}.  \cite{LR11} suggest that the monochromatic flux from a TDE may pass through three major evolutionary phases ($t^{-5/12}$, $t^{-5/3}$, and exponential decay) as it cools, depending upon the relation of the bandpass to the peak emission wavelength.  For X-rays, the $t^{-5/12}$ phase is not seen, whereas the $t^{-5/3}$ phase may last years.  \cite{Maksym13} suggest that there is some evidence for this late-time exponential decay.  Simulations by \cite{GR13} and \cite{GMR14} paint a more complicated picture, as $\dot{M}$ may be affected by the retention of an intact stellar core and may asymptote to $t^{-2.2}$.  

In order to evaluate the applicability of such decay models, we consider the plausible \Mbh\ range. \cite{Ghosh06} suggest $\Mbh\sim5\times10^4\;\Msun$ based on blackbody disk models, which requires sustained super-Eddington accretion in the 1990 epoch, given that $L_X\sim L_{Edd}$ implies $\Mbh\simgreat10^5\;\Msun$ for a bolometric correction of $\simgreat1.5$.  We consider a plausible upper bound for \Mbh,  given the $L_{bulge}-\Mbh$ relationship \citep[e.g.,][]{Marconi03,KH13}.  The applicability of $L_{bulge}-\Mbh$ in this regime is uncertain, and would be affected by the galaxy morphology \citep[see, e.g.][and references therein]{KH13}, including the nature of any nuclear component in SDSS J114726.69+494257.8.  Although \cite{Ghosh06} label this galaxy as a nucleated dwarf spheroidal, the Galaxy Zoo morphology is uncertain \citep{Lintott08}.  Since $L_{bulge}\simless L_{total}$, we find $\rm{log}(\Mbh/\Msun)\simless6.4$ for $I=15.98$ \citep{Ghosh06}, according to \cite{Jiang11}.

The 2009 observation of RBS 1032 is sufficiently late that it is likely to be in the \cite{LR11} exponential decay regime for $\Mbh\simgreat10^5\;\Msun$, which is compatible with the observations.  The {\it ROSAT} epochs may instead be described by a more gradual decay than $t^{-5/3}$ \citep[see][]{LR11}, if early accretion is significantly affected by stellar structure \citep{Lodato09} and $\dot{M}$ in the super-Eddington regime is strongly modulated by outflows \citep{DS10}.  For example, \cite{LR11} show that during the first $\sim4$ years of a disruption where $\Mbh=10^6\;\Msun$, $L_X$ may vary by an amount comparable to the RBS 1032 {\it ROSAT} data.  The super-Eddington phase may last $\sim0.76(\Mbh/10^7\;\Msun)^{-2/5}$ years \citep{Ulmer99}, or $\sim4.8$ years at $\Mbh\sim10^5\;\Msun$.  Alternately, a disrupted red giant may exhibit similar slowly-evolving $\dot{M}$ \citep{MGR12}.  When $n$ for $L_X \propto t^{-n}$ is left free, we find $n=2.5\pm1.3$, which is closer to the \cite{GR13} suggestion of $t^{-2.2}$ for a surviving stellar core than to the Keplerian $n=5/3$ value, but compatible with both.

\section{Discussion}

The small separation between 3XMM J114726.7+494257 and RBS 1032, and the lack of other likely \xmm\ counterparts to RBS 1032 both strongly suggest that they are the same X-ray source (see, e.g., the \citealt{Donato14} analysis of {\it EUVE} data for WINGS J1348).  If so, the \cite{Ghosh06} attribution of RBS 1032 to SDSS J114726.69+494257.8 is likely correct.  

As with other X-ray selected TDE candidates \citep[e.g.][]{Maksym10,Maksym13}, the extreme luminosity ($L_X\simgreat10^{43}\;$\es), supersoft spectrum $\Gamma\sim5$ at early times, and extreme variability (a factor of $\simgreat200$ decay) make this a strong candidate for a TDE.  This is not a surprising conclusion, given \cite{Wang04} suggest significantly elevated \tdr\ for nucleated dwarf spheroidal galaxies.  Other possible explanations can be rejected via arguments similar to those advanced by \cite{Maksym10,Maksym13} and others.  The X-ray spectrum is too soft for a GRB.  Supernovae may exhibit soft thermal spectra, but $L_X$ would make RBS 1032 among the most X-ray luminous supernovae \citep{Levan13}.  RBS 1032 is, however, significantly softer at early times than the super-luminous supernova SCP 06F6 ($\Gamma\sim5$ vs $\Gamma\sim2.5$).  The early X-ray emission in RBS 1032 is orders of magnitude too luminous for a stellar mass X-ray binary in SDSS J114726.69+494257.8.  \cite{Ghosh06} suggest a classical (and presumably Galactic) nova as possible but unlikely.  In addition to their arguments, we suggest the high galactic latitude ($b\sim64\degr$) makes a Galactic foreground object unlikely \citep{Maksym10,Maksym13}.  Classical novae in particular are rare at $b\simgreat30\degr$ \citep{IT12}.

\cite{Ghosh06} suggest that RBS 1032 may be an intermediate-mass black hole (IMBH) binary, possibly with a white dwarf secondary component.  At least one other IMBH binary has been suggested, ESO 243-09 HLX-1 \citep[][suggest a donor giant star]{Lasota11}.  As with ESO 243-09 HLX-1, an eccentric orbit could mimic the extreme variability seen in RBS 1032.  Additional X-ray monitoring observations could verify or disprove such an explanation.

Given that both Narrow-Line Seyfert 1 \citep[e.g.][]{Grupe95a} and Seyfert 2 \citep{Saxton11AGN} galaxies may exhibit extreme supersoft X-ray variability on timescales similarly short to TDEs, an AGN origin is typically the most difficult explanation to dismiss for an X-ray source with these characteristics.  Variability in such a case may be attributed to change in $\dot{M}$ or $N_H$.  \cite{Ghosh06} demonstrate, however, that RBS 1032 has no significant emission lines.  Steady accretion by an AGN is therefore disfavored.  TDEs may temporarily excite emission lines \citep[e.g.][]{BogdanovicEtAl04,CE11,Gezari12}, but such lines might not be visible at $t_0\simgreat15$ years.

The \xmm\ spectrum is harder than previous epochs ($\Gamma\sim3$ vs $\sim5$).  Such evolution has been seen in X-ray observations at $t_0\simgreat10\;$years \citep{Halpern04}.  This could be a state change due to a lower $\dot{M}$, or a weak ``hard" physical component only visible when the soft component subsides.  The lack of emission lines indicates little star formation or nuclear activity. Contamination due to hot gas or X-ray binaries is thus likely to small ($\simless \rm{few}\times10^{39}\;$\es).

Given the importance of {\it RASS} for determining \tdr, \cite{Donley02} remains an important reference for discussion of the observed \tdr.  At $\tdr\sim10^{-5}\;\tdru$, this also remains one the most conservative \tdr\ values in the literature.  The discovery of single {\it RASS} flare would not have a major impact on \cite{Donley02} even given they only assume 3 outbursts from inactive galaxies.  However, the presence of other unidentified flares (undetected due to long periods of shallow evolution, for example) might require a re-evaluation of \cite{Donley02} and its applicability to TDEs.  \cite{Donley02} assume that any flare which varies by $\simgreat20$ during {\it RASS} and the preceding $\sim$six months would be detected by their study.  This assumption was reasonable according to TDE theory at the time, and was confirmed by all known TDEs identified by {\it ROSAT}.

RBS 1032, though observed by {\it ROSAT} three times over $\sim4$ years, only varied by $\sim5$ and thus was not selected.  Under certain models, however, TDEs may have relatively shallow X-ray light curves for years at a time.  In particular, certain models \citep[e.g.][]{LR11} may have relatively shallow X-ray light curves for years.  But even with modest deviations from best-fit $t^{-5/3}$ decay, RBS 1032 could fall within the \cite{Donley02} window.  As a result, we expect that not all {\it RASS} TDEs had been detected.  The TDE in Abell 3571 identified by \cite{Cappelluti09} also complicates the  \cite{Donley02} rate estimation, as the {\it RASS} $F_X$ upper limit falls earlier than their best-fit $t_0$, but is also above their earliest detected X-ray flux.  Given the flare's proximity to the diffuse intracluster X-ray emission from Abell 3571, and the likely importance of galaxy clusters in TDE production \citep{Wang04,Cappelluti09,Maksym10,Maksym13}, such cluster background may be important to any {\it RASS}-derived \tdr.

Given this state of affairs, we therefore argue that the results of \cite{Donley02} should be considered a lower limit to \tdr, and that long baseline X-ray studies (such as may use {\it RASS} and \cha, \xmm, {\it Swift} or {\it eROSITA}) are necessary to investigate the occurrence of TDEs which may display such gradual evolution.  This assertion is supported by the recent work of \cite{KS14}, who found several {\it ROSAT} flares in combination with the \xmm\ archive \citep[as predicted by][]{Donley02}.  Observations to follow up bright {\it ROSAT} sources may yield additional TDEs.  Also, long-baseline studies of TDE candidates from other systematic surveys \citep[such as the three as-yet-unfollowed candidates from][]{Esquej07} may prove helpful in better determining \tdr\ and the long-term X-ray behavior of TDE light curves.

%TDE behaviour is clearly far from well-understood, however, and many more candidates are needed to investigate the full range of associated observational phenomena.

\section{Conclusions}

By comparing archival \xmm\ observations with previous work by \cite{Ghosh06}, we confirm that RBS 1032 is associated with the dwarf galaxy SDSS J114726.69+494257.8.  We suggest that RBS 1032 is also likely to be a TDE.  Its peak luminosity ($L_X\simgreat10^{43}\;$\es), softness ($\Gamma\sim5$), and extreme variability (factor of $\sim200$) in a quiescent galaxy are typical of X-ray selected TDEs.  Via the absence of AGN emission lines, \cite{Ghosh06} have already demonstrated that an AGN is an unlikely explanation.  Other explanations, such as GRBs, supernovae, and X-ray binaries are disfavored.  Although the nuclear black hole could be an IMBH, a more massive black hole ($\simgreat 10^6\;$\Msun) would be consistent with the host galaxy luminosity, depending upon its morphology.  Higher-resolution images, such as with the {\it Hubble Space Telescope}, would allow more thorough analysis of the host galaxy's morphology and properties.

We have not ruled out an explanation similar to \cite{Ghosh06}, that RBS 1032 may result from an IMBH binary.  If so, the extreme variability may imply an eccentric orbit, as suggested for ESO 243-09 HLX-1 \citep{Lasota11}.  Additional X-ray monitoring observations would constrain this scenario.  If RBS 1032 remains weak or undetected, the TDE scenario would be favored.

The discovery of a new TDE candidate in archival {\it ROSAT} All-Sky Survey data supports the idea that, as per \cite{Donley02}, additional TDEs may yet be discovered in {\it RASS} via X-ray follow up of bright {\it ROSAT} sources, and in comparison with \cha\ and \xmm\ archives.  RBS 1032 also implies that the variability constraint of $\simgreat20$ may leave some X-ray selected TDEs undiscovered, such that the outburst rate in \cite{Donley02} should be taken as a lower limit for the true tidal disruption rate.  Further {\it RASS} follow-up may help determine whether recent disruption rates of $\sim5-10\times10^{-5}\;\tdru$ \citep[e.g.][]{Maksym10,Maksym13,Esquej08} are more reasonable.  This is consistent with more recent models of multi-band TDE light curve variability \citep[e.g.][]{LR11}.  Recent work by \cite{KS14} using archival {\it ROSAT} and \xmm\ data already points to a higher rate ($\sim3\times10^{-5}\;\tdru$).  More generally, targeted X-ray follow-up of otherwise inactive galaxies which demonstrate bright X-ray activity may be fruitful in exploring the full range of TDE phenomena.
%% XXX COMMENT ON KS14



%% Included in this acknowledgments section are examples of the
%% AASTeX hypertext markup commands. Use \url without the optional [HREF]
%% argument when you want to print the url directly in the text. Otherwise,
%% use either \url or \anchor, with the HREF as the first argument and the
%% text to be printed in the second.

\acknowledgments
We thank the anonymous referee, whose comments significantly improved this paper.

WPM, JAI and DL acknowledge support from the University of Alabama, and from NASA ADAP Grant NNX10AE15G.  WPM acknowledges support from a University of Alabama Research Stimulus Program grant.

This research has made use of the NASA/IPAC Infrared Science Archive, operated by the Jet Propulsion Laboratory, California Institute of Technology, under contract with the National Aeronautics and Space Administration.

Funding for SDSS-III has been provided by the Alfred P. Sloan Foundation, the Participating Institutions, the National Science Foundation, and the U.S. Department of Energy Office of Science. The SDSS-III web site is http://www.sdss3.org/.

SDSS-III is managed by the Astrophysical Research Consortium for the Participating Institutions of the SDSS-III Collaboration.

{\it Facilities:} \facility{XMM-NEWTON}, \facility{ROSAT}, \facility{SDSS}.


\begin{thebibliography}{49}
\expandafter\ifx\csname natexlab\endcsname\relax\def\natexlab#1{#1}\fi

\bibitem[{{Ahn} {et~al.}(2014){Ahn}, {Alexandroff}, {Allende Prieto}, {Anders},
  {Anderson}, {Anderton}, {Andrews}, {Aubourg}, {Bailey}, {Bastien}, \&
  et~al.}]{SDSSdr10}
{Ahn}, C.~P., {et~al.} 2014, \apjs, 211, 17

%\bibitem[{{Arnaud}(1996)}]{Arnaud96}
%{Arnaud}, K.~A. 1996, in Astronomical Society of the Pacific Conference Series,
%  Vol. 101, Astronomical Data Analysis Software and Systems V, ed.
%  {G.~H.~Jacoby \& J.~Barnes}, 17--20

\bibitem[{{Bade} {et~al.}(1996){Bade}, {Komossa}, \& {Dahlem}}]{BKD96}
{Bade}, N., {Komossa}, S., \& {Dahlem}, M. 1996, \aap, 309, L35

\bibitem[{{Bloom} {et~al.}(2011){Bloom}, {Giannios}, {Metzger}, {Cenko},
  {Perley}, {Butler}, {Tanvir}, {Levan}, {O'Brien}, {Strubbe}, {De Colle},
  {Ramirez-Ruiz}, {Lee}, {Nayakshin}, {Quataert}, {King}, {Cucchiara},
  {Guillochon}, {Bower}, {Fruchter}, {Morgan}, \& {van der Horst}}]{Bloom11}
{Bloom}, J.~S., {et~al.} 2011, Science, 333, 203

\bibitem[{{Bogdanovi{\' c}} {et~al.}(2004){Bogdanovi{\' c}}, {Eracleous},
  {Mahadevan}, {Sigurdsson}, \& {Laguna}}]{BogdanovicEtAl04}
{Bogdanovi{\' c}}, T., {Eracleous}, M., {Mahadevan}, S., {Sigurdsson}, S., \&
  {Laguna}, P. 2004, ApJ, 610, 707

\bibitem[{{Burrows} {et~al.}(2011){Burrows}, {Kennea}, {Ghisellini}, {Mangano},
  {Zhang}, {Page}, {Eracleous}, {Romano}, {Sakamoto}, {Falcone}, {Osborne},
  {Campana}, {Beardmore}, {Breeveld}, {Chester}, {Corbet}, {Covino},
  {Cummings}, {D'Avanzo}, {D'Elia}, {Esposito}, {Evans}, {Fugazza}, {Gelbord},
  {Hiroi}, {Holland}, {Huang}, {Im}, {Israel}, {Jeon}, {Jeon}, {Jun}, {Kawai},
  {Kim}, {Krimm}, {Marshall}, {P.~M{\'e}sz{\'a}ros}, {Negoro}, {Omodei},
  {Park}, {Perkins}, {Sugizaki}, {Sung}, {Tagliaferri}, {Troja}, {Ueda},
  {Urata}, {Usui}, {Antonelli}, {Barthelmy}, {Cusumano}, {Giommi}, {Melandri},
  {Perri}, {Racusin}, {Sbarufatti}, {Siegel}, \& {Gehrels}}]{Burrows11}
{Burrows}, D.~N., {et~al.} 2011, \nat, 476, 421

\bibitem[{{Cappelluti} {et~al.}(2009){Cappelluti}, {Ajello}, {Rebusco},
  {Komossa}, {Bongiorno}, {Clemens}, {Salvato}, {Esquej}, {Aldcroft},
  {Greiner}, \& {Quintana}}]{Cappelluti09}
{Cappelluti}, N., {et~al.} 2009, \aap, 495, L9

\bibitem[{{Cash}(1979)}]{Cash79}
{Cash}, W. 1979, \apj, 228, 939

\bibitem[{{Clausen} \& {Eracleous}(2011)}]{CE11}
{Clausen}, D., \& {Eracleous}, M. 2011, \apj, 726, 34

\bibitem[{{Dickey} \& {Lockman}(1990)}]{Dickey90}
{Dickey}, J.~M., \& {Lockman}, F.~J. 1990, \araa, 28, 215

\bibitem[{{Donato} {et~al.}(2014){Donato}, {Cenko}, {Covino}, {Troja},
  {Pursimo}, {Cheung}, {Fox}, {Kutyrev}, {Campana}, {Fugazza}, {Landt}, \&
  {Butler}}]{Donato14}
{Donato}, D., {et~al.} 2014, \apj, 781, 59

\bibitem[{{Donley} {et~al.}(2002){Donley}, {Brandt}, {Eracleous}, \&
  {Boller}}]{Donley02}
{Donley}, J.~L., {Brandt}, W.~N., {Eracleous}, M., \& {Boller}, T. 2002, \aj,
  124, 1308

\bibitem[{{Dotan} \& {Shaviv}(2010)}]{DS10}
{Dotan}, C., \& {Shaviv}, N.~J. 2010, arXiv:1004.1797

\bibitem[Esquej et~al.(2007)]{Esquej07} 
{Esquej}, P., {Saxton}, R.~D., {Freyberg}, M.~J., {Read}, A.~M., {Altieri}, B., 
{Sanchez-Portal}, M., \& {Hasinger}, G., 2007, \aap, 462, L49 

\bibitem[{{Esquej} {et~al.}(2008){Esquej}, {Saxton}, {Komossa}, {Read},
  {Freyberg}, {Hasinger}, {Garc{\'{\i}}a-Hern{\'a}ndez}, {Lu}, {Zaur{\'{\i}}n},
  {S{\'a}nchez-Portal}, \& {Zhou}}]{Esquej08}
{Esquej}, P., {et~al.} 2008, \aap, 489, 543

\bibitem[{{Fischer} {et~al.}(1998){Fischer}, {Hasinger}, {Schwope}, {Brunner},
  {Boller}, {Tr{\"u}mper}, {Voges}, \& {Neizvestnyj}}]{RBS1}
{Fischer}, J.-U., {Hasinger}, G., {Schwope}, A.~D., {Brunner}, H., {Boller},
  T., {Tr{\"u}mper}, J., {Voges}, W., \& {Neizvestnyj}, S. 1998, Astron.
  Nachr., 319, 347

\bibitem[{{Gezari} {et~al.}(2012){Gezari}, {Chornock}, {Rest}, {Huber},
  {Forster}, {Berger}, {Challis}, {Neill}, {Martin}, {Heckman}, {Lawrence},
  {Norman}, {Narayan}, {Foley}, {Marion}, {Scolnic}, {Chomiuk}, {Soderberg},
  {Smith}, {Kirshner}, {Riess}, {Smartt}, {Stubbs}, {Tonry}, {Wood-Vasey},
  {Burgett}, {Chambers}, {Grav}, {Heasley}, {Kaiser}, {Kudritzki}, {Magnier},
  {Morgan}, \& {Price}}]{Gezari12}
{Gezari}, S., {et~al.} 2012, \nat, 485, 217

\bibitem[{{Ghosh} {et~al.}(2006){Ghosh}, {Suleymanov}, {Bikmaev}, {Shimansky},
  \& {Sakhibullin}}]{Ghosh06}
{Ghosh}, K.~K., {Suleymanov}, V., {Bikmaev}, I., {Shimansky}, S., \&
  {Sakhibullin}, N. 2006, \mnras, 371, 1587

\bibitem[{{Grupe} {et~al.}(1995){Grupe}, {Beuerman}, {Mannheim}, {Thomas},
  {Fink}, \& {de Martino}}]{Grupe95a}
{Grupe}, D., {Beuerman}, K., {Mannheim}, K., {Thomas}, H., {Fink}, H.~H., \&
  {de Martino}, D. 1995, \aap, 300, L21+

\bibitem[{{Guillochon} {et~al.}(2014){Guillochon}, {Manukian}, \&
  {Ramirez-Ruiz}}]{GMR14}
{Guillochon}, J., {Manukian}, H., \& {Ramirez-Ruiz}, E. 2014, \apj, 783, 23

\bibitem[{{Guillochon} \& {Ramirez-Ruiz}(2013)}]{GR13}
{Guillochon}, J., \& {Ramirez-Ruiz}, E. 2013, \apj, 767, 25

\bibitem[{{Halpern} {et~al.}(2004){Halpern}, {Gezari}, \&
  {Komossa}}]{Halpern04}
{Halpern}, J.~P., {Gezari}, S., \& {Komossa}, S. 2004, \apj, 604, 572

\bibitem[{{Hills}(1975)}]{Hills75}
{Hills}, J.~G. 1975, Nature, 254, 295

\bibitem[{{Imamura} \& {Tanabe}(2012)}]{IT12}
{Imamura}, K., \& {Tanabe}, K. 2012, \pasj, 64, 120

\bibitem[{{Irwin} {et~al.}(2010){Irwin}, {Brink}, {Bregman}, \&
  {Roberts}}]{Irwin10}
{Irwin}, J.~A., {Brink}, T.~G., {Bregman}, J.~N., \& {Roberts}, T.~P. 2010,
  \apjl, 712, L1

\bibitem[{{Jiang} {et~al.}(2011){Jiang}, {Greene}, \& {Ho}}]{Jiang11}
{Jiang}, Y.-F., {Greene}, J.~E., \& {Ho}, L.~C. 2011, \apjl, 737, L45

\bibitem[Khabibullin \& Sazonov(2014)]{KS14} 
{Khabibullin}, I., \& {Sazonov}, S.\ 2014, \mnras, accepted, arXiv:1407.6284 

\bibitem[{{Komossa} \& {Bade}(1999)}]{KB99}
{Komossa}, S., \& {Bade}, N. 1999, \aap, 343, 775

\bibitem[{{Komossa} \& {Greiner}(1999)}]{KG99}
{Komossa}, S., \& {Greiner}, J. 1999, \aap, 349, L45

\bibitem[Komossa et~al.(2004)]{Komossa04} 
{Komossa}, S., {Halpern}, J., {Schartel}, N., {Hasinger}, G., {Santos-Lleo}, M.,
\& {Predehl}, P.  2004, \apjl, 603, L17 

\bibitem[{{Kormendy} \& {Ho}(2013)}]{KH13}
{Kormendy}, J., \& {Ho}, L.~C. 2013, \araa, 51, 511

\bibitem[{{Lasota} {et~al.}(2011){Lasota}, {Alexander}, {Dubus}, {Barret},
  {Farrell}, {Gehrels}, {Godet}, \& {Webb}}]{Lasota11}
{Lasota}, J.-P., {Alexander}, T., {Dubus}, G., {Barret}, D., {Farrell}, S.,
  {Gehrels}, N., {Godet}, O., \& {Webb}, N. 2011, \apj, 735, 89

\bibitem[{{Levan} {et~al.}(2013){Levan}, {Read}, {Metzger}, {Wheatley}, \&
  {Tanvir}}]{Levan13}
{Levan}, A.~J., {Read}, A.~M., {Metzger}, B.~D., {Wheatley}, P.~J., \&
  {Tanvir}, N.~R. 2013, \apj, 771, 136

\bibitem[{{Lin} {et~al.}(2011){Lin}, {Carrasco}, {Grupe}, {Webb}, {Barret}, \&
  {Farrell}}]{Lin11}
{Lin}, D., {Carrasco}, E.~R., {Grupe}, D., {Webb}, N.~A., {Barret}, D., \&
  {Farrell}, S.~A. 2011, \apj, 738, 52

\bibitem[{{Lintott} {et~al.}(2008){Lintott}, {Schawinski}, {Slosar}, {Land},
  {Bamford}, {Thomas}, {Raddick}, {Nichol}, {Szalay}, {Andreescu}, {Murray}, \&
  {Vandenberg}}]{Lintott08}
{Lintott}, C.~J., {et~al.} 2008, \mnras, 389, 1179

\bibitem[{{Lodato} {et~al.}(2009){Lodato}, {King}, \& {Pringle}}]{Lodato09}
{Lodato}, G., {King}, A., \& {Pringle}, J. 2009, \mnras, 392, 332

\bibitem[{{Lodato} \& {Rossi}(2011)}]{LR11}
{Lodato}, G., \& {Rossi}, E.~M. 2011, \mnras, 410, 359

\bibitem[{{MacLeod} {et~al.}(2012){MacLeod}, {Guillochon}, \&
  {Ramirez-Ruiz}}]{MGR12}
{MacLeod}, M., {Guillochon}, J., \& {Ramirez-Ruiz}, E. 2012, \apj, 757, 134

\bibitem[{{Maksym} {et~al.}(2010){Maksym}, {Ulmer}, \& {Eracleous}}]{Maksym10}
{Maksym}, W.~P., {Ulmer}, M.~P., \& {Eracleous}, M. 2010, \apj, 722, 1035

\bibitem[{{Maksym} {et~al.}(2013){Maksym}, {Ulmer}, {Eracleous}, {Guennou}, \&
  {Ho}}]{Maksym13}
{Maksym}, W.~P., {Ulmer}, M.~P., {Eracleous}, M.~C., {Guennou}, L., \& {Ho},
  L.~C. 2013, \mnras, 435, 1904

\bibitem[{{Maksym} {et~al.}(2014)}]{Maksym14}
{Maksym}, W.~P., {Ulmer}, M., {Roth}, K., {Irwin}, J., {Dupke}, R., {Ho}, L., {Keel}, W., 
\& {Adami}, C. 2014, \mnras, accepted, arXiv:1407.6737

\bibitem[{{Marconi} \& {Hunt}(2003)}]{Marconi03}
{Marconi}, A., \& {Hunt}, L.~K. 2003, \apjl, 589, L21

\bibitem[{{Rees}(1988)}]{Rees88}
{Rees}, M.~J. 1988, Nature, 333, 523

\bibitem[{{Saxton} {et~al.}(2011){Saxton}, {Read}, {Esquej}, {Miniutti}, \&
  {Alvarez}}]{Saxton11AGN}
{Saxton}, R., {Read}, A., {Esquej}, P., {Miniutti}, G., \& {Alvarez}, E. 2011,
  in Narrow-Line Seyfert 1 Galaxies and their Place in the Universe

\bibitem[{{Saxton} {et~al.}(2012){Saxton}, {Read}, {Esquej}, {Komossa},
  {Dougherty}, {Rodriguez-Pascual}, \& {Barrado}}]{Saxton12}
{Saxton}, R.~D., {Read}, A.~M., {Esquej}, P., {Komossa}, S., {Dougherty}, S.,
  {Rodriguez-Pascual}, P., \& {Barrado}, D. 2012, \aap, 541, A106

%\bibitem[{{Schwope} {et~al.}(2000){Schwope}, {Hasinger}, {Lehmann}, {Schwarz},
%  {Brunner}, {Neizvestny}, {Ugryumov}, {Balega}, {Tr{\"u}mper}, \&
%  {Voges}}]{RBS2}
%{Schwope}, A., {et~al.} 2000, Astrnon. Nachr., 321, 1

\bibitem[{{Shcherbakov} {et~al.}(2013){Shcherbakov}, {Pe'er}, {Reynolds},
  {Haas}, {Bode}, \& {Laguna}}]{Shcherbakov13}
{Shcherbakov}, R.~V., {Pe'er}, A., {Reynolds}, C.~S., {Haas}, R., {Bode}, T.,
  \& {Laguna}, P. 2013, \apj, 769, 85

\bibitem[{{Ulmer}(1999)}]{Ulmer99}
{Ulmer}, A. 1999, ApJ, 514, 180

\bibitem[{{Voges} {et~al.}(1999){Voges}, {Aschenbach}, {Boller},
  {Br{\"a}uninger}, {Briel}, {Burkert}, {Dennerl}, {Englhauser}, {Gruber},
  {Haberl}, {Hartner}, {Hasinger}, {K{\"u}rster}, {Pfeffermann}, {Pietsch},
  {Predehl}, {Rosso}, {Schmitt}, {Tr{\"u}mper}, \& {Zimmermann}}]{RASS}
{Voges}, W., {et~al.} 1999, \aap, 349, 389

\bibitem[{{Wang} \& {Merritt}(2004)}]{Wang04}
{Wang}, J., \& {Merritt}, D. 2004, \apj, 600, 149

%\bibitem[{{Wright}(2006)}]{Wright06}
%{Wright}, E.~L. 2006, \pasp, 118, 1711

\bibitem[{{XMM-Newton Survey Science Centre}(2013)}]{3XMM}
{XMM-Newton Survey Science Centre}. 2013, VizieR Online Data Catalog, 9044, 0

\bibitem[{{Zickgraf} {et~al.}(2003){Zickgraf}, {Engels}, {Hagen}, {Reimers}, \&
  {Voges}}]{Zickgraf03}
{Zickgraf}, F.-J., {Engels}, D., {Hagen}, H.-J., {Reimers}, D., \& {Voges}, W.
  2003, \aap, 406, 535

\end{thebibliography}
\end{document}